\documentclass[USenglish]{article}	

\usepackage[utf8]{inputenc}				
\usepackage[big,online]{dgruyter}	
\usepackage{lmodern} 
\usepackage{microtype}
\usepackage[numbers,square,sort&compress]{natbib}

\theoremstyle{dgthm}

\theoremstyle{dgdef}

\begin{document}

	\articletype{Research Article}
	\received{Month	DD, YYYY}
	\revised{Month	DD, YYYY}
  \accepted{Month	DD, YYYY}
  \journalname{De~Gruyter~Journal}
  \journalyear{YYYY}
  \journalvolume{XX}
  \journalissue{X}
  \startpage{1}
  \aop
  \DOI{10.1515/sample-YYYY-XXXX}

\title{Bound states in the continuum in periodic structures with structural disorder}
\runningtitle{BIC in disordered structures}

\author[1]{Ekaterina E. Maslova}
\author[1,2]{Mikhail V. Rybin}
\author[1,2]{Andrey A. Bogdanov$^\dagger$} 
\author*[1]{Zarina F. Sadrieva} 
\affil[1]{\protect\raggedright 
School of Physics and Engineering, ITMO University, 197101, St. Petersburg, Russia, e-mail: $^\dagger$a.bogdanov@metalab.ifmo.ru; *z.sadrieva@metalab.ifmo.ru}
\affil[2]{\protect\raggedright 
Ioffe Institute, 194021,  St. Petersburg, Russia}
	
	
\abstract{
We study the effect of structural disorder on the transition from the bound states in the continuum (BICs) to quasi-BICs by the example of the periodic photonic structure composed of two layers of parallel dielectric rods. We uncover the specificity in the robustness of the symmetry-protected and accidental BICs against various types of structural disorder. We analyze how the spatial mode localization induced by the structural disorder results in an effective reduction of the system length and limits the Q factor of quasi-BICs. Our results are essential for the practical implementation of BICs especially in natural and self-assembled photonic structures, where the structural disorder plays a crucial role.
}


\maketitle

\section{Introduction} 

Bound states in the continuum (BICs) are the specific solutions of the wave equation which are spatially localized despite the fact that their frequencies lay in the radiation continuum. Initially, BICs were proposed in quantum mechanics by von Neumann and Wigner~\cite{Neumann}. In the last two decades, BICs have been observed in many photonic structures such as photonic crystals, metasurfaces,\cite{hsu2016bound,koshelev2018meta,han2018all}, chains of coupled resonators~\cite{bulgakov2017bound,bulgakov2015light,sadrieva2019experimental,sidorenko2021observation}, waveguides~\cite{zou2015guiding,yu2019photonic,bulgakov2008bound,bezus2018bound,bykov2020bound} and even single subwavelength resonators supporting quasi-BIC in a form of supercavity modes~\cite{rybin2017PRL,PhysRevLett.121.033903,odit2021observation}.  
A divergent radiative quality (Q) factor, strong spatial localization, and drastic enhancement of the incident field make BICs very prospective for many applications including lasers~\cite{kodigala2017lasing,Wang2020,cui2018multiple,midya2018coherent,Song2020,mohamed2020topological,hwang2021ultralow,wu2020room,ha2018directional}, optical filters~\cite{foley2014symmetry,foley2015normal,cui2016normal}, biological and chemical sensors~\cite{romano2018surface, romano2018label, srivastava2019terahertz, chen2020toroidal, maksimov2020refractive}, and non-linear photonics~\cite{kivshar2017meta,koshelev2019nonlinear,krasikov2018nonlinear,carletti2018giant,bulgakov2019nonlinear,kravtsov2020nonlinear,carletti2019high,anthur2020continuous,maksimov2020optical,koshelev2020subwavelength}. 

A genuine BIC with an infinitely large radiative Q factor is a mathematical idealization. In real samples, the radiative Q factor becomes finite due to roughness, diffraction on the edges and into the substrate, and other imperfections inevitably appearing during the fabrication. Thus, in practice, a genuine BIC turns into a {\it quasi-BIC} (q-BIC) manifesting itself in the scattering spectra as a high-Q resonance~\cite{timofeev2018optical,taghizadeh2017quasi,sadrieva2017transition,koshelev2018asymmetric}. The material absorption also results in a decrease in the total Q factor of q-BIC and makes it hardly recognizable in the experimental scattering spectra. The identification of the dominant loss mechanism is the crucial point allowing to enhance the performance of the photonic devices based on BICs. 

Most of the systems supporting q-BIC are periodic (metasurfaces, grating, chains etc). The {\it structural disorder} in such systems drastically affects their optical properties resulting in non-trivial Fano resonance evolution, light localization, coherent back-scattering etc  ~\cite{wolf1985weak,wiersma1997localization,poddubny2012fano,limonov2012optical,liu2019disorder}. The disorder effects are most essential in self-assembled and natural photonic structures~\cite{galisteo2011self,astratov2002interplay,fan1995theoretical}. Despite the important role of the structural disorder in periodic photonic structures, it is weakly studied in application to BICs and their transition to q-BICs. In particular, the coupled-wave theory (CWT) framework was developed to study the Q factor of symmetry-protected BIC in dielectric gratings with randomly perturbed filling factor and the element position~\cite{Ni:17}. The radiation losses of symmetry-protected BICs were shown to decrease quadratically with the fluctuation amplitude. It was shown in Ref.~\cite{xiao2018band} that disorder can results in the formation of BIC band in multi-chain and multi-layer systems. The robustness of symmetry-protected BICs was partly analyzed in Ref.~\cite{chen2018nearly} in the framework of the Fano-Anderson model applied to the system of parallel coupled waveguide.

Here we provide a rigorous analysis of the Q factor of both symmetry-protected and accidental BICs in a two-layered periodic array of infinitely long dielectric rods accounting for the structural disorder in the position of the rods keeping them parallel, i.e. preserving the translation symmetry. In contrast to the Ref.~\cite{Ni:17}, we uncover two concurrent radiation loss mechanisms due to a finite number of periods and scattering due to the structural disorder. We find the disorder amplitude up to which the vertical and horizontal fluctuations in the rod position contribute to the radiation losses independently. We demonstrate that accidental BIC is more resistant to the fluctuation of the position along the direction of periodicity rather than to the perturbation of distance between two layers of rods. Contrary, the symmetry-protected BIC shows greater sensitivity to the fluctuation of the distance between the layers. Also, we reveal the disorder-induced spatial localization of the q-BICs and prove that it limits the Q factor of q-BICs if the array of rods is longer than the localization length.

\begin{figure}[t]\centering
	\includegraphics[width = 0.6\linewidth ]{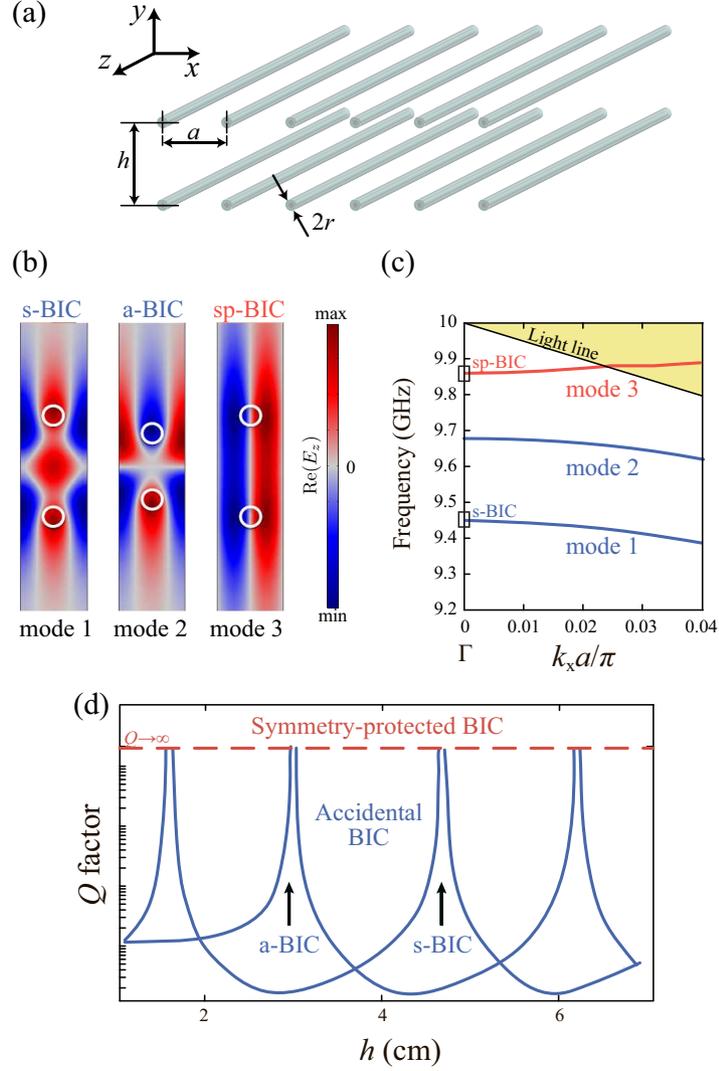}%
	\caption{
		\label{fig:fig1_infinite}
    Bound states in the continuum in two-layered structure of infinite dielectric rods. (a) Schematic view of the photonic structure. (b) Electric field $E_z$ distribution of accidental (mode 1, mode 2) and symmetry-protected (mode 3) BICs in the $\Gamma$-point. The mode profiles are plotted for different distances between layers $h$ ($h=4.7$~cm for modes 1 and 3; $h=3$~cm for mode 2). (c) The band diagram of TE-polarized modes. (d) Q factor versus distance between layers $h$. }
    \end{figure}
    
\section{Results and discussions}

\subsection{Infinite array without structural disorder}

We start the analysis from the consideration of the infinite periodic structure without structural disorder shown in Fig.~\ref{fig:fig1_infinite}(a). It consists of two identical periodic arrays of infinitely long dielectric rods. The radius of the rods $r=0.5~$cm and the period $a=3~$cm. Thus, the unit cell of the structure contains two rods spaced apart by distant $h$. For further analysis we will use permittivity of the rods $\varepsilon=2.1$ that corresponds to low-loss polymers, for example, Teflon in the frequency range 0.1-10~GHz~\cite{jin2006terahertz,haas1976}. The advantage of such low-refractive-index materials is that the electromagnetic field is mostly localized outside the rods, i.e. in air, where the losses are negligible. Thus, the materials absorption is suppresses substantially and we will not account for it in the further analysis. The considered two-layer system is convenient as it provides a simple way to precisely tune the accidental and symmetry-protected BICs~\cite{ndangali2010electromagnetic,ndangali2013resonant,bulgakov2018propagating,hemmati2019resonant,shuai2013double,shuai2013coupled,marinica2008bound}. 

According to the Bloch theorem, the electric field of eigenmodes can be written as $\mathbf{E}(x,y,z)=\mathbf{U}_{s,k_x}(x,y)e^{ik_xx+ik_zz}$. Here, $k_z$ is the wavenumber corresponding to the direction of the translation symmetry, $k_x$ is the Bloch wavenumber, $s$ is the index of the band and we will omit for simplicity, $\mathbf{U}_{s,k_x}(x,y)$ is the periodic function of $x$. The BICs in such system appear only if either $k_z=0$ or $k_x=0$. Further, we will limit our analysis to the case of TE-polarized modes [$\mathbf{E}=(0,0,E_z)$] with $k_z=0$. For TE modes with $k_z=0$, we can write $E_{z}(x,y) = U_{k_x}(x,y) e^{i k_x x}$. The periodic Bloch amplitude can be expanded into the Fourier series as follows:
\begin{equation}
     U_{k_x}(x,y) = \sum_n C_{n,k_x}(y) e^{i\frac{2\pi n}{a}x},
\end{equation}
where $n$ is the index of the diffraction channel.  At the frequencies above the light line ($\omega/c>|k_x|$), the mode leaks from the structure to the radiation continuum via the open diffraction channels. BIC appears when the leakage into all open diffraction channels is forbidden, i.e. the complex Fourier coefficients $C_{n,k_x}(y)$ -- the amplitudes of the outgoing waves  -- are zero. In the subwavelength regime $\lambda<a$, only the zeroth diffraction channel is open. Thus, the amplitude of the outgoing leaky wave is defined by the zeroth Fourier coefficient $C_{0,k_x}(y)$. By the definition, it is equal to the field averaged over the period, i.e. $C_{0,k_x}(y) = \langle U_{0,k_x}(x,y)\rangle_x$. For the structures having the time-reversal and $\pi$-rotational symmetries, denoted as $TC_{2}^{y}$, the coefficient $C_{0,k_x}(y)$ becomes pure real for BICs~\cite{hsu2013observation,zhen2014topological,bulgakov2017topological}. At the $\Gamma$-point (the center of the Brillouin zone), $U_{k_x}(x,y)$ is either odd or even function of $x$ since the photonic structure is $C_2^y$-invariant~\cite{Sakoda,Ivchenko1995}. Obviously, for an odd function $U_{k_x}(x,y)$ the zero-order Fourier coefficient vanishes. In this case,  the \emph{symmetry-protected} BIC occurs -- the coupling to the radiation continuum disappear due to the point-symmetry of the structure. In the case of the even mode, the spatial average $\langle U_{0,k_x}(x,y)\rangle_x$ may vanish not only due to the symmetry reasons but at specific values of the structure parameters such as period, the radius of rods, permittivity, interlayer distance resulting in the appearance of \emph{accidental} BIC also called \emph{tunable} BIC.

Figures~\ref{fig:fig1_infinite}(b) and \ref{fig:fig1_infinite}(c) show the spectra of TE-modes and the electric field distribution for possible BICs appeared in the $\Gamma$-point. Mode~3 is odd with respect to  $C_{2}^{y}$ transformation and, thus, it has a symmetry-protected BIC labeled as sp-BIC. Modes~1 and 2 are even with respect to $C_{2}^{y}$ transformation and, thus, they can have only accidental BICs. The dependence of the Q factor for all three modes in the $\Gamma$-point vs the distance between the layers $h$ is shown in Fig.~\ref{fig:fig1_infinite}(d). One can see from this figure that the BICs in the $\Gamma$-point of modes 1 and 2 appear only at the specific distances between the layers, while the Q factor of the sp-BIC is insensitive to $h$. The spectral position of the $\Gamma$-point for all three modes vs $h$ is shown in Fig.~\ref{fig:ap1} in Appendix~\ref{appendix}. Modes 1 and 2 have different symmetries with respect to the $y\rightarrow -y$ transformation. Mode 1 is even (symmetric) and mode 2 is odd (anti-symmetric). Thus, we denote the corresponding BICs as s-BIC and a-BIC.  These BICs can be considered as Fabry-Perot resonances formed due to the total reflection from the array of the rods~\cite{karagodsky2011,Karagodsky2012,bulgakov2017bound}.

 \begin{figure}[t]\centering
	\includegraphics[width = 0.7\linewidth ]{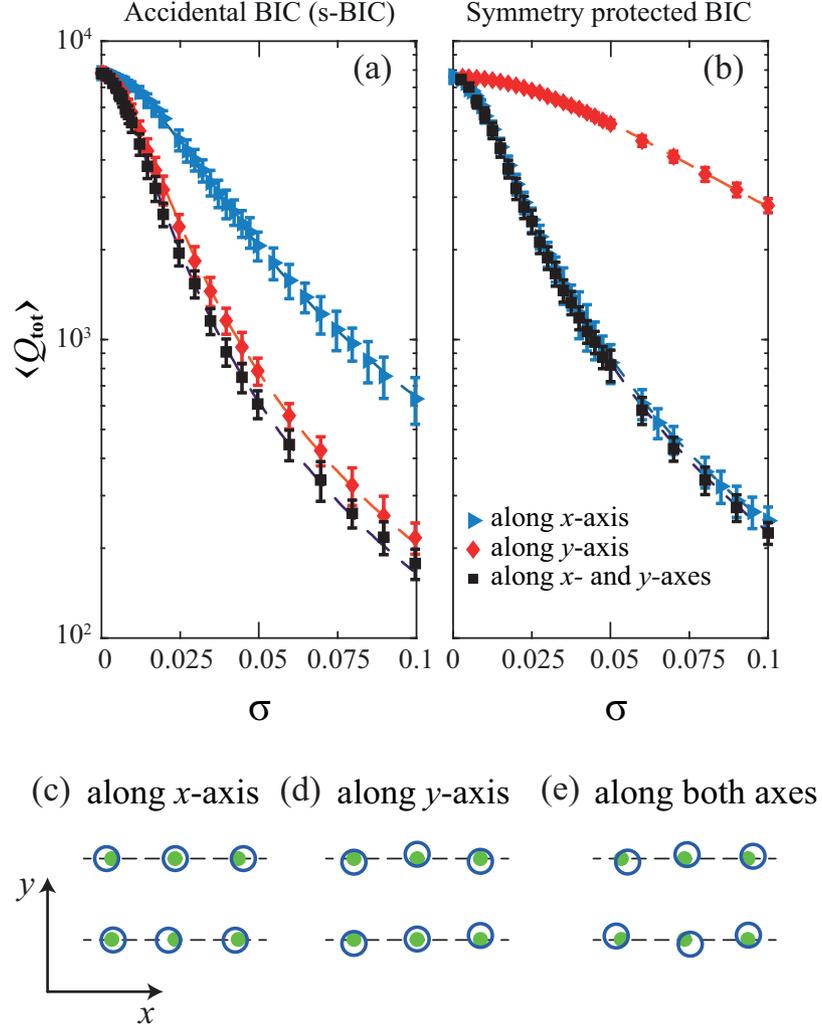}
	\caption{
		\label{fig:disorders}
      Dependence of average $Q_\text{tot}$ factor of disordered structures on disorder amplitude $\sigma$ for (d) accidental s-BIC and (e) symmetry-protected sp-BIC. Bars indicate the standard deviation. Number of periods $N=100$. Schematic view of the rod array with the structural disorder (c) along $x$-axis, (d) along $y$-axis and (e) along both axes.  Green solid circles show the ordered structure, blue open circles show the structures with disorder.}
    \end{figure}

        \begin{figure}[t]\centering
	\includegraphics[width = 0.8\linewidth ] {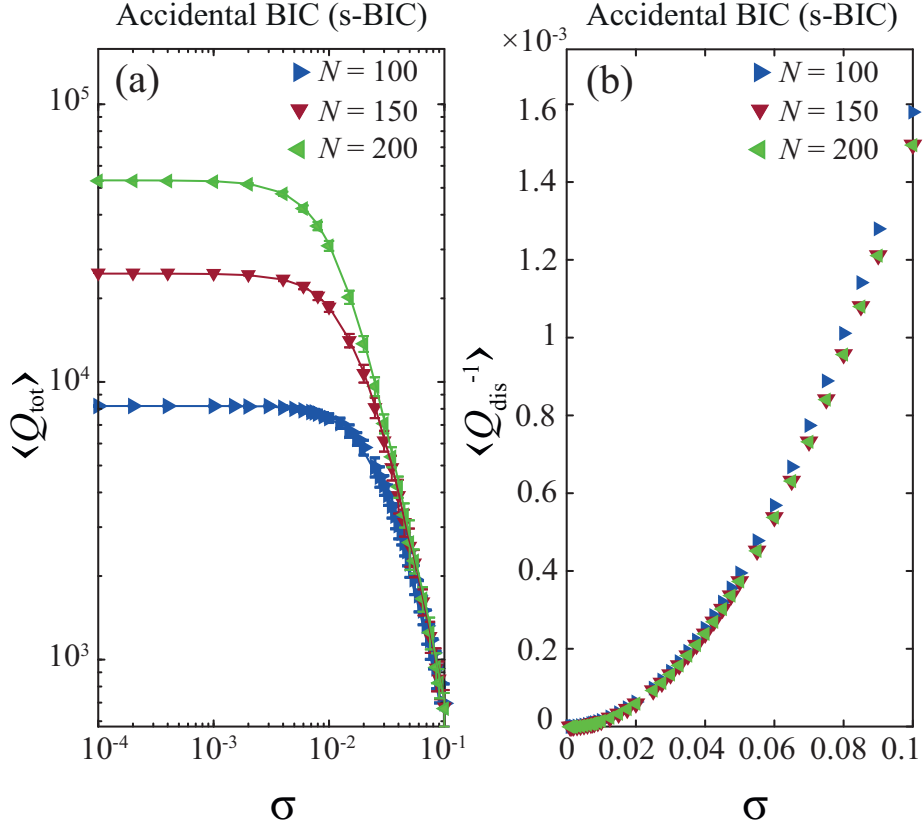}
	\caption{
		\label{fig:approx}
   (a) Average $Q_\text{tot}$ factor of the accidental s-BIC supported by the structure with disorder along $x$ axis with different number of periods $N$. Vertical bars denote the standard deviation. Solid lines show the approximation. (b) Average $Q^{-1}_{\text{dis}}$ of the accidental s-BIC supported by the structure with disorder along $x$ axis with different number of periods $N$. Blue, red and green symbols correspond to $N = 100, 150, 200$, respectively.
   }
    \end{figure}

\subsection{Structural disorder and Q factor}    
    
In this section, we consider the structure of finite size, consisting of $N$ periods, accounting for the structural disorder. In this case, the total Q factor is contributed by two parts
\begin{equation}
    1/Q_\text{tot}=1/Q_\text{ord} + 1/Q_{\text{dis}}
    \label{eq:ord_plus_sis}
\end{equation}
 as we neglect absorption and the effect of surface roughness. Here, $Q_{\text{dis}}$ is responsible for the radiation due to the structural disorder; $Q_\text{ord}$ is responsible for the radiation due to the finite size of the structure and it is a function of $N$. Since the finite structure is not true periodic, the diffraction directions are smeared out resulting in additional diffraction losses. Therefore, a genuine BIC transforms into a resonant state (quasi-BIC) with finite $Q$ factor~\cite{sadrieva2018experimental,sadrieva2019experimental}. Recently, the $Q$ factor of a symmetry-protected quasi-BIC was shown to grow as $\sim N^2$, see Ref.~\cite{bulgakov2017lightenh}, while the $Q$ factor of an accidental quasi-BIC emerging at the center of the Brillouin zone increases as $\sim N^3$, see Refs.~\cite{bulgakov2017lightenh,bulgakov2019high}. The dependence of $Q_\text{ord}$ on $N$ calculated numerically for the structure under consideration is shown in Table~\ref{tab:Qord_vs_N}.

\begin{table}[b]\centering
\caption{\label{tab:Qord_vs_N} 
Radiative Q factor ($Q_\text{ord}$) for different types of BICs in structures of finite size consisting of $N$ periods. 
}
\begin{tabular}{lrrr}
Number of periods, $N$ 	& 100           & 150          & 200          \\ \midrule
sp-BIC          		&   7600    &   22500    &   47300  \\ 
a-BIC           		&   10300    &   33900    &   78800  \\
s-BIC           		&   7800    &   24700    &   60200  \\ 
\end{tabular}
\end{table}

As a next step, we account for the contribution of the structural disorder $Q_{\text{dis}}$ to the total quality factor $Q_\text{tot}$. We will consider a model of the uncorrelated disorder introducing a fluctuation to the position of the rod keeping all the rods parallel. To gain deeper insight into how the structural disorder affects symmetry-protected and accidental BICs, we first consider the fluctuations in the rod positions along the $x$ and $y$ axes independently and then account for these shifts simultaneously. In both cases, the shift of the rod is given by $\delta t =  R \cdot \sigma \cdot a$. Here, $a$ is the period of the structure, $\sigma$ is a fluctuation amplitude (disorder amplitude); $R$ is a uniformly distributed random number in the range $[-1;~1]$. 


To analyze $Q_\text{dis}$ we use the finite element method calculating the complex eigenfrequencies and total Q factors of the disordered array of rods with the finite number of periods $N$. We collect the data from ensembles of 100 structures for each value of amplitude $\sigma$ at the given number of periods $N$. Then we average the calculated Q factors over the ensemble as $\langle Q_\text{tot} \rangle = \sum_{i=1}^{m} Q_\text{tot}^i/ m $, where $m$ is the number of ensembles.

Figure~\ref{fig:disorders} exhibits decreasing of the Q factor for both symmetry-protected and accidental BICs (sp-BIC and s-BIC) with the growing amplitude of disorder $\sigma$. As it was mentioned above, the at-$\Gamma$ accidental BICs (a-BIC and s-BIC) require particular values of the distance between layers $h$ at which the round-trip phase shift accumulates to a value being multiple of $2\pi$. Therefore, the accidental BICs are more sensitive to the fluctuations in rod position along the $y$ axis rather than along the $x$ axis [see Fig.~\ref{fig:disorders}(a)]. Thus, when we account the shift of rods along both coordinates, we observe that the random shifts along the $y$-axis make the main contribution to the decreasing of the Q factor for the accidental BICs.  
In contrast, the symmetry-protected BIC (sp-BIC) exist at any distance between the layers (Fig.~\ref{fig:fig1_infinite}). Therefore, the $y$-axis disorder does not affect much its Q factor. However, the random shifts of rods along the $x$-axis induce an additional polarization (dipole moment) along the $x$-axis resulting in radiative losses. Thus, the symmetry-protected BIC is most sensitive to the lateral shifts of meta-atoms inside the unit cell [Fig.~\ref{fig:disorders}(b)].

Next, we calculate the dependence of the total Q factor ($Q_\text{tot}$) of the accidental s-BIC on the disorder amplitude $\sigma$ for three structures of different length ($N=100,150,200$) accounting for random shifts of the rods only along the $x$-axis [Fig.~\ref{fig:approx}a]. One can see from this figure
that the Q factor becomes sensitive to structural disorder only when the disorder amplitude reaches a threshold $\sigma\sim0.01$. For smaller $\sigma$, the Q factor is completely defined by the radiation losses due to the finite size of the structure.  In contrast, for $\sigma > 0.01$ the main contribution to the radiation losses is made by the scattering events on the structural disorder resulting in quadratic decaying of the Q factor with $\sigma$. This quadratic dependence is observed for both types of the structural disorder (along the $x$ and $y$ axes) (see Fig.~\ref{fig:ap2} in Appendix~\ref{appendix2}). The similar dependence of the $Q_{\text{dis}}$ factor was shown recently for eigenmodes of the periodic dielectric grating composed of supercells with fluctuations in grooves positions and filling factor~\cite{Ni:17}.

       \begin{figure}[t]\centering
	\includegraphics[width = 0.7\linewidth ]{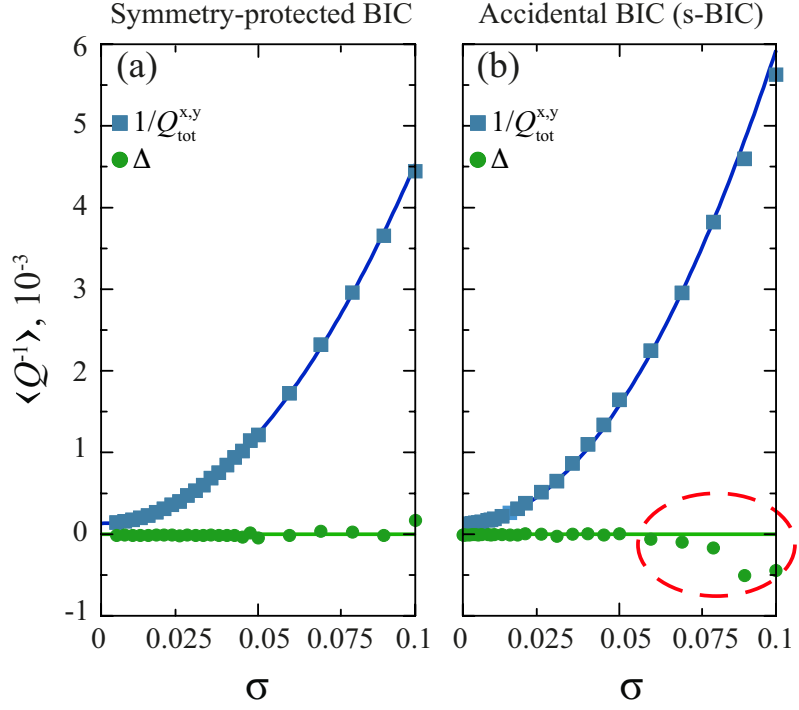}
	\caption{
		\label{fig:sum}
    Inverse average $Q$ factor: (a) symmetry-protected BIC, (b) accidental s-BIC. Blue squares show dependence of inverse average $Q$ factor on the disorder degree along both axes, green points are the difference according to Eq.~\eqref{eq:delta}. Solid lines are guides for an eye only. Red dashed circle shows the region of $\sigma$, where Eq.~\eqref{eq:delta} is not valid.}
    \end{figure}

\begin{figure}[h!]\centering
	\includegraphics[width = 0.6\linewidth ]{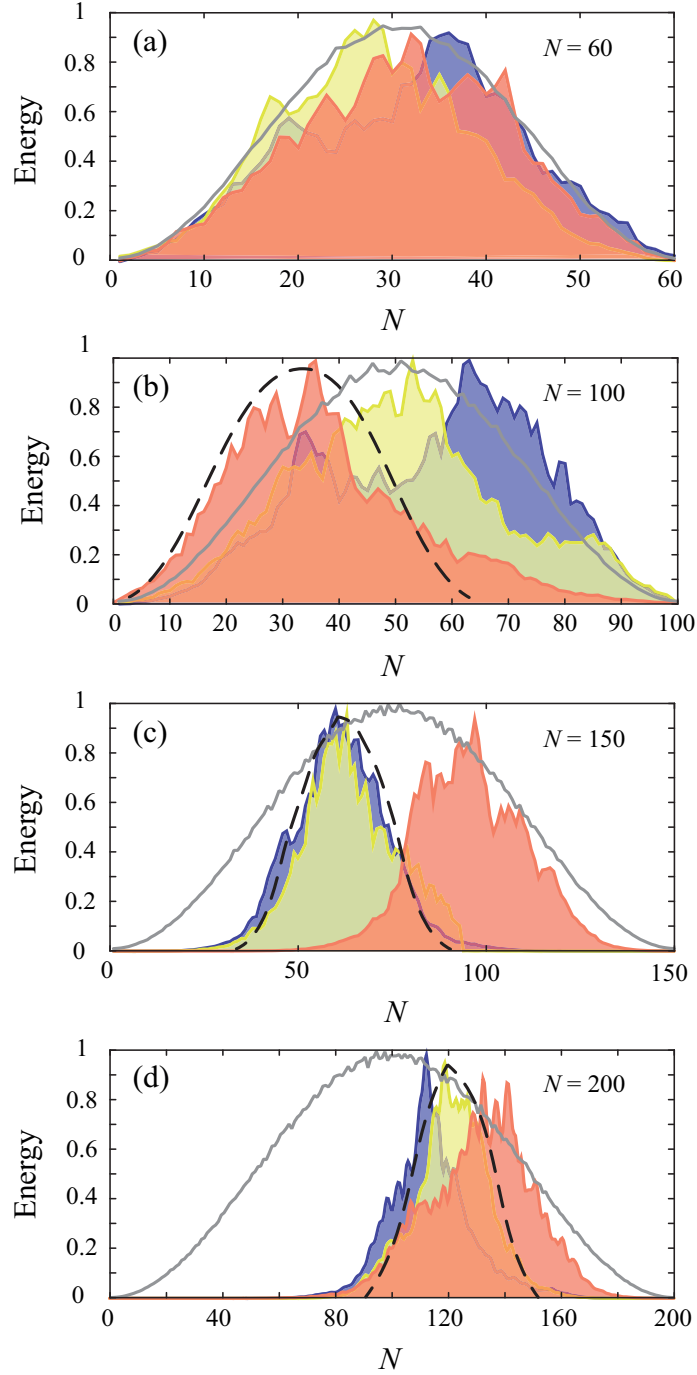}%
	\caption{
		\label{fig:localiz}
Electromagnetic energy distribution along chain at the accidental BIC (s-BIC). Curves with shaded area with different colors show several realization of the disorder $\sigma = 0.085$ along $x$ axis. Grey curves are the distribution for the structures with no disorder. (a) $N = 60$, (b) $N = 100$, (c) $N = 150$, (d) $N = 200$. The black dashed curves in panels (b)-(d) show the energy distribution in the chain of $N=60$ with no disorder $\sigma=0$ for comparison.}
    \end{figure}

Next,  we examine whether the losses due to disorder along the $x$ and $y$ axes can be decomposed into two independent terms corresponding to the disorder along each axis separately 
\begin{equation}
    \frac{1}{Q_{\text{dis}}^{x,y}} = \frac{1}{Q_{\text{dis}}^{x}} + \frac{1}{Q_{\text{dis}}^{y}}.
    \label{eq:sum}
\end{equation}
Usually, it is true only when the radiative losses caused due to both of these shifts are small. To analyze the validity of this decomposition, we consider an auxiliary quantity characterizing the deviation from it
\begin{equation}
    \Delta = \frac{1}{Q_{\text{dis}}^{x,y}} - \frac{1}{Q_{\text{dis}}^{x}} - \frac{1}{Q_{\text{dis}}^{y}}.
    \label{eq:delta}
\end{equation}
If the composition \eqref{eq:sum} is correct then $\Delta$ has to be zero. Figure~\ref{fig:sum} demonstrates the dependence of $1/Q_{\text{dis}}^{x,y}$ and $\Delta$ on an amplitude of disorder $\sigma$ for the symmetry-protected (panel a) and accidental s-BIC (panel b). As it can be seen from the figure, decomposition \eqref{eq:sum} breaks for the accidental s-BIC for $\sigma>0.7$. For the symmetry-protected BIC, there are some deviations of $\Delta$ for large $\sigma$ but they substantially smaller than ones for the accidental s-BIC. It is worth mentioning that large disorder amplitudes can result in the spatial localization of the mode at a scale less than the length of the structure, thus, reducing its effective length.

\subsection{Effect of spatial localization}

As we discussed previously, the Q factor of all types of BICs depends on the size of the sample. From other hand it is known that disorder can result in spatial localization of wavefunctions (eigenmodes) that can be considered as effective reduction of the sample size~\cite{Lifshits1988,evers2008anderson,billy2008direct,segev2013anderson,poddubny2012fano}. In this section, we analyze how the structural disorder results in spatial localization of BICs and how it affects their Q factors.

As we discussed in the previous section, there is a deviation from Eq.~\ref{eq:sum} for large disorder amplitudes for the accidental BIC (s-BIC). We suppose that this deviation is related to the wave localization inherited to one-dimension systems with structural disorder~\cite{Lifshits1988,poddubny2012fano}, which leads to an effective reduction of the structure length. We examine the amplitude of the electric field distribution along the structures of different length. 

Figure~\ref{fig:localiz} shows the envelop curves (mode profiles) of the accidental BIC (s-BIC) for the structure of various length ($N=100,150,200$) for the fixed disorder amplitude $\sigma=0.085$. The shaded areas of different colours correspond to different realization from the ensemble. The grey curves corresponds to the envelope function of the accidental s-BIC in structures without disorder. To estimate the localization length, we superpose the envelope function of the accidental s-BIC in structure with $N=60$ (dashed curve). One can see that for disorder amplitude $\sigma=0.085$ the wavefunction is localized on around 60 periods. 
From one hand, the structural disorder localizes the wavefunction inside the structure, thus, the loses due radiation from the ends of the chain are negligible. Therefore, this loss mechanism is essential as for regular structure. From other hand, the structural disorder reduces the effective length of the structure limiting the Q factor. The localization can be neglected for the most practical shorter structures, it has to be taken into account for larger, self-assembled, and natural structures, though.

\section{Conclusion}

In conclusion, we have uncovered how the uncorrelated structural disorder affects the symmetry-protected and accidental at-$\Gamma$ BICs in the one-dimensional periodic structure composed of two layers of dielectric rods. Regardless of the direction along which the position of rods are perturbed, the Q factor of both symmetry-protected and accidental BICs decays quadratically with the amplitude of the disorder $\sigma$. Importantly, the symmetry-protected BIC is more resistant to the fluctuation of the distance between layers (along $y$) rather than the fluctuation of position along $x$. And vice versa, the accidental BIC is more robust against the fluctuation along the direction of periodicity (along $x$). We have shown that the fluctuations of position along the direction of periodicity and distance between the layers contribute to the losses independently for low disorder amplitudes. For the large values of disorder amplitude $\sigma$ or long structures, the localization of electromagnetic energy suppresses the total Q factor due to the effective reduction of the system length. We anticipate our findings provide useful guidelines for practical implementation of resonators supporting bound states in the continuum in self-assembled and natural structures, where the role of structural disorder is the most relevant.

\section*{APPENDIX: BIC's frequency in a periodic structure without disorder}
\label{appendix}
\begin{figure}[h]\centering
	\includegraphics[width = 0.66\linewidth ]{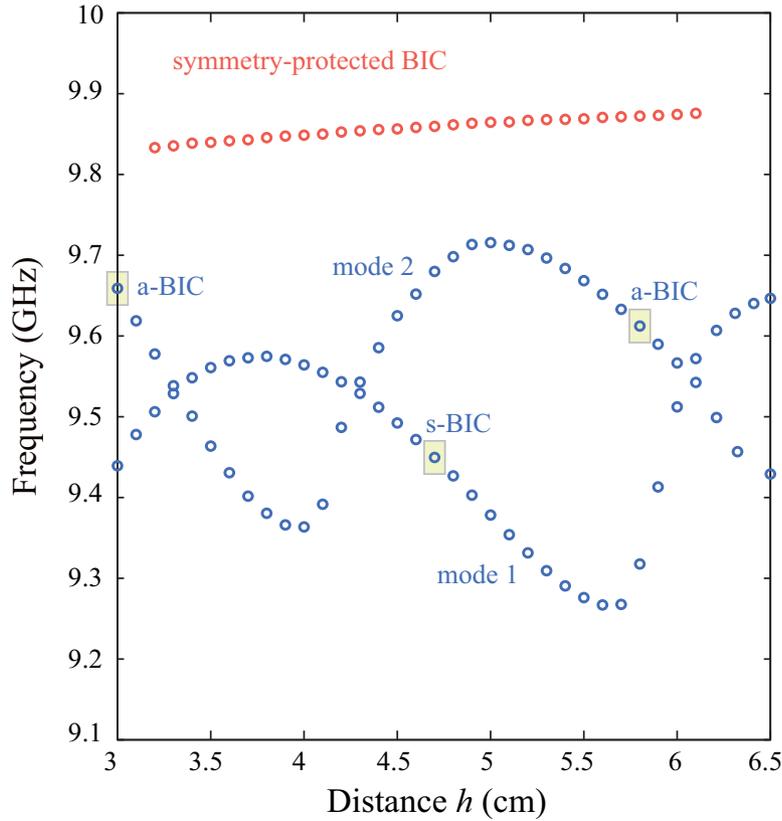}%
	\caption{
		\label{fig:ap1}
    Frequency of three lowest modes of the structure a=under consideration. The symmetry-protected BIC is denoted as sp-BIC, symmetric accidental BIC -- as s-BIC, asymmetric accidental BIC -- as a-BIC. The yellow rectangles correspond to accidental BICs [see Fig.~\ref{fig:fig1_infinite}(b)].}
    \end{figure}

  Figure~\ref{fig:ap1} shows how the frequencies of sp-, a- and s-BICs depend on the distance between layers $h$ of the infinite periodic structure.
 While the frequency of the sp-BIC increases slowly, the frequencies of the mode 1 and mode 2, at which the accidental BICs occur, exhibit complicate behaviour.
  
\subsection*{Radiative losses due to structural disorder }\label{appendix2}
     \begin{figure}[h!]\centering
	\includegraphics[width = 0.7\linewidth ]{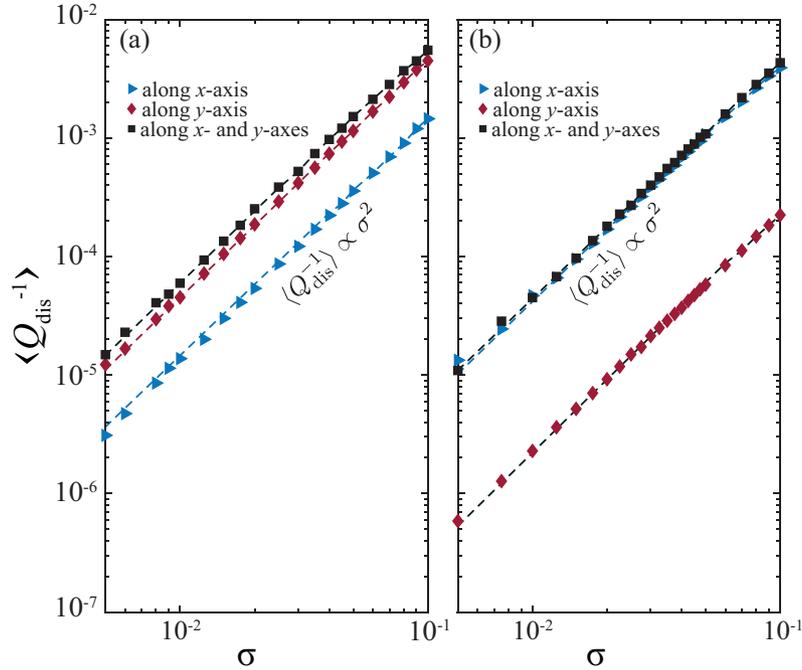}%
	\caption{
		\label{fig:ap2}
    Average $Q^{-1}_\text{dis}$ factor of (a) the accidental BIC and (b) symmetry-protected BIC supported by the structure with disorder. Dashed lines show quadratic approximation.}
    \end{figure}
 Figure~\ref{fig:ap2} shows that, for both symmetry-protected and accidental BICs, the losses due to various types of structural disorder exhibit quadratic dependence on the disorder amplitude.

\begin{funding}
This work was supported by the Russian Science Foundation (grant no. 21-19-00677). A.B. acknowledges the BASIS foundation and the Council on Grants of the
President of the Russian Federation (MK-2224.2020.2).
\end{funding}


\end{document}